\documentclass[11pt]{article}

\usepackage[utf8]{inputenc}
\usepackage[T1]{fontenc}
\usepackage[margin=1in]{geometry}
\usepackage{times}
\usepackage{microtype}
\usepackage{amsmath,amssymb}
\usepackage{booktabs}
\usepackage{enumitem}
\usepackage{titlesec}
\usepackage{authblk}
\usepackage[hidelinks]{hyperref}
\usepackage{xcolor}
\usepackage{abstract}
\usepackage[numbers,sort&compress]{natbib}
\usepackage{fancyhdr}
\usepackage{parskip}

\titleformat{\section}{\normalfont\large\bfseries}{\thesection}{0.6em}{}
\titleformat{\subsection}{\normalfont\normalsize\bfseries}{\thesubsection}{0.6em}{}
\titlespacing*{\section}{0pt}{1.4em}{0.6em}

\title{\bfseries Anticipatory Data Governance in the Age of AI:\\[0.3em]
\large Emerging Signals in Data Access, Reuse, and Sovereignty}

\author[1]{Adam Zable}
\author[1,2]{Stefaan Verhulst}
\affil[1]{The Governance Lab (The GovLab), New York University}
\affil[2]{The Data Tank}

\date{June 2026}

\begin{document}
\maketitle
\thispagestyle{fancy}

\begin{abstract}
\noindent
Artificial intelligence is transforming how data is collected, accessed, shared, governed, and reused, and in doing so is unsettling the assumptions on which two decades of open-data policy were built. Prevailing governance frameworks presuppose that reuse is largely visible and traceable, that institutions retain meaningful control over information flows, and that humans are the primary consumers of data. Generative and inferential AI systems violate each of these premises: they absorb data into opaque training pipelines, generate sensitive knowledge without accessing underlying datasets, and increasingly act as the dominant producers and consumers of information.

This paper reports findings from a structured participatory foresight study comprising two expert forecasting studios convened by The GovLab between 2025 and 2026. The studios brought together nineteen senior practitioners spanning official statistics, digital and trade policy, open science, AI governance, geospatial systems, and public-sector innovation across multiple jurisdictions. Applying a qualitative signal-scanning methodology grounded in the horizon-scanning and anticipatory-governance traditions, we elicited, clustered, and thematically synthesized emerging developments in data access, governance, and reuse, and stress-tested them against practitioner experience.

We identify seven convergent signals: (1) the open-data paradigm is under strain; (2) data ecosystems are becoming machine-centric and AI-mediated; (3) inference is reshaping the foundations of data governance; (4) data infrastructure is becoming harder to sustain; (5) governance is fragmenting across institutions and jurisdictions; (6) sovereignty and security are driving a turn toward strategic control; and (7) data-sharing models require stronger incentives and benefit-sharing mechanisms. We further map the reinforcing feedback loops that couple these signals, showing how interventions in one domain propagate risks and opportunities across the wider ecosystem. We argue that data governance is becoming inseparable from AI governance, digital public infrastructure, economic strategy, democratic resilience, and geopolitical competition, and we outline an agenda for \emph{anticipatory} data governance capable of adapting before dependencies, risks, and missed opportunities become locked in. The contribution is diagnostic rather than predictive: the signals offer an evidence-informed framework for reasoning about structural shifts already underway, not a forecast of specific technological outcomes.

\vspace{0.6em}
\end{abstract}

\section{Introduction}
The world's relationship to digital data is changing rapidly. AI tools have generated significant excitement for their potential to unlock new insights, expand access to knowledge, create new forms of economic and social value, and help solve public problems \citep{verhulst2019leveraging,cgc2023}. At the same time, they have intensified longstanding debates around access, ownership, attribution, privacy, labor rights, security, extraction, and the responsible reuse of public-interest data \citep{taylor2025,verhulst2025reimagining,longpre2024}. Questions that once sat at the margins of data policy have moved to the center.

Governments, international organizations, civil society groups, and companies are responding with a growing range of governance approaches, from summits, executive orders, and industry initiatives to international agreements, laws, regulatory frameworks, and new institutional models \citep{roberts2024}. Yet it remains difficult to distinguish short-term developments from deeper structural shifts. While discussion often focuses on the latest AI breakthrough, the more consequential question for policymakers is how these technologies are reshaping the systems, assumptions, and governance models that determine how data is accessed, shared, and used.

To examine these changes, The GovLab convened two forecasting studios bringing together experts working in data governance, digital policy, open science, AI governance, and public-sector innovation. Participants explored emerging trends in data access, governance, and reuse; examined the forces driving those trends; and considered what they may mean for the future of open data and public-interest data ecosystems.

The studios were designed to support three linked aims: recognizing emerging signals, building the capacity to interpret and respond to them, and informing the design of more anticipatory data policy. This paper contributes to that effort by identifying major shifts affecting data access, openness, reuse, infrastructure, sovereignty, and trust, while also surfacing the governance questions and institutional capacities that will shape the future of data policy. The discussions identified seven signals that point to significant changes already underway. These signals suggest that the future of data governance will be shaped by advances in AI alongside evolving expectations around trust, stewardship, infrastructure, sovereignty, reciprocity, and public value.

The remainder of the paper is organized as follows. Section~\ref{sec:method} details the study's methodology, including participant selection, elicitation protocol, and the analytic procedure used to derive and validate the signals, together with an explicit statement of the approach's limitations. Section~\ref{sec:signals} presents the seven signals. Section~\ref{sec:implications} analyzes the cross-cutting themes and reinforcing feedback loops that connect them and develops the case for anticipatory data governance. Section~\ref{sec:conclusion} concludes.

\section{Methodology}
\label{sec:method}

\subsection{Research design and rationale}
This study employs a qualitative, participatory foresight design situated within the horizon-scanning and anticipatory-governance traditions \citep{fuerth2009,ramos2014,wilkinson2017}. Foresight methods are appropriate when the object of inquiry is characterized by deep uncertainty, rapid technological change, and an absence of settled historical data from which to extrapolate---conditions that characterize the interaction between AI and data governance. Rather than attempting to predict discrete future events, the design seeks to surface \emph{weak signals}: early, often ambiguous indicators of structural change that are observable in the present but whose systemic implications are not yet fully understood \citep{fuerth2009}. This orientation follows established practice in strategic foresight, in which the analytic goal is to expand the range of plausible futures under consideration and to strengthen institutional capacity to detect and respond to change, rather than to assign probabilities to specific outcomes \citep{wilkinson2017,kimbell2019}. It also extends a growing body of work on \emph{data-driven anticipatory governance}, which reconceives foresight as a designed system of actors, processes, and technologies rather than a one-off exercise \citep{maffei2020,marcucci2025}.

We selected an expert-elicitation format---the ``forecasting studio''---as the primary instrument. The studio format combines structured group deliberation with facilitated sense-making, and is well suited to eliciting tacit, cross-domain knowledge that is not yet documented in the published literature. It also reflects a broader methodological convergence in which quantitative forecasting and qualitative foresight are increasingly blended within a single anticipatory toolkit \citep{marcucci2025}. Because the developments of interest are emergent and cut across legal, technical, economic, and institutional domains, no single documentary source or disciplinary vantage point provides adequate coverage; convening diverse expert practitioners allows distributed, frontline observations to be pooled, contrasted, and synthesized.

\subsection{Participant selection}
Participants were recruited through purposive, criterion-based sampling. The sampling frame targeted senior practitioners and researchers with direct, decision-level experience of data governance in one or more of the following domains: official statistics, open data and open science, digital and data policy, AI governance, geospatial data systems, digital trade, and public-sector digital innovation. Selection criteria prioritized (i)~demonstrated seniority and operational responsibility; (ii)~diversity of institutional vantage point, spanning national governments, multilateral and regional organizations, standards bodies, private-sector platforms, civil-society organizations, and academic institutions; and (iii)~geographic diversity, with deliberate inclusion of perspectives from the Global South to counteract the over-representation of high-income jurisdictions common in this literature.

The final cohort comprised nineteen participants. Institutional affiliations included national governments (e.g., Taiwan, Brazil, Germany), multilateral and intergovernmental bodies (the World Bank, the OECD, the Publications Office of the European Union, the ASEAN Secretariat), standards and open-knowledge organizations (the Open Data Charter, the Open Geospatial Consortium, Creative Commons), private technology firms (Microsoft, Hugging Face), and research institutes across several continents. The full roster of participants and affiliations is provided in Appendix~\ref{app:participants}. This composition was intended to maximize the diversity of observed signals rather than to constitute a statistically representative sample; the study makes no claim to representativeness in the survey-research sense.

\subsection{Data collection: the forecasting studios}
Two forecasting studios were convened. Each studio followed a common facilitated protocol structured in three phases:

\begin{enumerate}[leftmargin=1.4em,itemsep=0.2em]
    \item \textbf{Signal elicitation.} Participants were prompted to identify developments in data access, governance, and reuse that they judged to be emerging, under-recognized, or structurally significant. Elicitation was deliberately open-ended to avoid anchoring participants to a predetermined framework and to allow unanticipated signals to surface.
    \item \textbf{Driver analysis.} For each candidate signal, participants examined the underlying forces---technological, economic, institutional, political, and social---driving the development, and interrogated whether it represented a transient phenomenon or a durable structural shift.
    \item \textbf{Implications and stress-testing.} Participants collectively assessed the governance implications of each signal and stress-tested it against their own operational experience, identifying supporting evidence, counter-examples, and boundary conditions. Concrete cases volunteered by participants (for instance, the inference of protected traffic-signal timing from aggregated navigation traces) were used to ground and probe abstract claims.
\end{enumerate}

Discussions were documented through detailed facilitator notes and rapporteur records. The deliberative format allowed claims to be immediately contested and refined by other domain experts, providing a form of real-time peer scrutiny that a series of isolated interviews would not.

\subsection{Analysis: signal synthesis}
Analysis proceeded through iterative thematic synthesis. Raw elicited material was first inventoried and coded to identify recurrent developments, concerns, and framings. Codes recurring across both studios and across multiple institutional vantage points were retained as candidate signals, on the principle that convergence among independently situated experts strengthens confidence that a development is systemic rather than idiosyncratic. Candidate signals were then clustered by underlying mechanism and consolidated: overlapping or nested items were merged, and items that reflected a single institution's circumstances without wider resonance were set aside. This process yielded the seven signals reported in Section~\ref{sec:signals}.

In a final analytic step, we examined the relationships \emph{between} signals, mapping the reinforcing and countervailing feedback loops through which a shift in one domain conditions outcomes in others. This relational analysis, reported in Section~\ref{sec:implications}, treats the signals not as an independent list but as an interacting system.

\subsection{Limitations}
Several limitations follow from the design and should frame interpretation of the findings. First, the study is qualitative and interpretive; the signals represent an evidence-informed reading of emerging developments, not quantified trends or forecasts, and no probabilities are attached to them. Second, although participant selection was deliberately diverse, a nineteen-person expert cohort is not statistically representative, and the synthesis inevitably reflects the vantage points present in the room; domains and regions less represented among participants may be correspondingly under-weighted. Third, expert elicitation is subject to well-documented cognitive and social biases, including deference to high-status participants and convergence on salient contemporary concerns; the facilitated, multi-domain format and the stress-testing phase were intended to mitigate but cannot fully eliminate these effects. Fourth, as a point-in-time exercise conducted in 2025--2026, the study captures a fast-moving landscape at a particular moment. We regard these constraints as acceptable given the study's diagnostic aim---to structure reasoning about structural change under uncertainty---rather than a predictive one.

\section{Seven Emerging Signals}
\label{sec:signals}

\subsection{Signal 1: The Open-Data Paradigm Is Under Strain}
One theme that surfaced repeatedly across the studios was the growing tension between traditional open-data models and the realities of the AI era. Open-data policies were built in part on the assumption that data reuse would be relatively visible and understandable. Institutions could generally see who was using their data, for what purposes, and with what outcomes. Generative AI challenges that logic. Data can now be absorbed into training pipelines as model inputs, learned patterns, or generated outputs, making downstream use difficult to trace, attribute, audit, or contest.

The category itself is also becoming too broad. ``Open data'' now encompasses government statistics, scientific datasets, legal texts, geospatial data, scraped web content, community-contributed data, synthetic data, and machine-generated data. These forms differ significantly in their users, risks, governance needs, and pathways to public value, yet they are often treated under the same policy umbrella.

This is especially problematic for non-personal, anonymized, collective, and community-level data, where governance frameworks remain underdeveloped despite rising strategic and commercial value. Such data may fall outside personal-data protection regimes, yet still generate significant risks when combined by AI systems to produce new inferences or commercial value.

These developments are creating a growing legitimacy challenge. Data shared for transparency, research, public service, civic participation, or collective benefit may later be incorporated into commercial AI systems with limited visibility, accountability, or return to the original providers. Legal and licensing frameworks designed for human-scale sharing, including Creative Commons-style approaches, often struggle to preserve reciprocity when reuse occurs at machine scale. As a result, openness is increasingly experienced by some creators, institutions, and communities as exposure to extraction, raising the risk that contributors may respond by restricting access to critical information. In response, policymakers are increasingly exploring more differentiated approaches to access that account for context, purpose, trust, reciprocity, and public value.

\subsection{Signal 2: Data Ecosystems Are Becoming Machine-Centric and AI-Mediated}
A second theme centered on the changing relationship between humans, machines, and data. Data ecosystems were largely designed for people to browse, interpret, and use information. Increasingly, however, machines are becoming the primary consumers and producers of data. In statistical systems, legal repositories, and public data infrastructures, APIs, bots, AI systems, and automated pipelines account for a growing share of activity. At the same time, sensors, satellites, vehicles, cameras, and infrastructure systems generate vast volumes of machine-native data, often processed at the edge before it ever reaches a centralized system or human reviewer.

AI is also changing the types of data that matter. Structured, tabular datasets are now joined by text, images, audio, video, geospatial information, synthetic data, and other unstructured or multimodal forms that can be consumed, combined, and repurposed at scale. Combined with developments such as geospatial-AI convergence, world models, and synthetic data, these shifts move the governance challenge from managing relatively static datasets to overseeing high-volume, real-time data flows that are harder to observe, interpret, and audit.

The meaning of access is shifting as a result. Data may be technically public but effectively unusable for machines if it lacks metadata, documentation, provenance, or machine-readable formats. Conversely, information originally intended for human audiences can be harvested, recombined, and operationalized at machine scale in ways institutions never anticipated. Automated traffic and AI scraping can overwhelm legacy infrastructure, while poorly documented or inconsistently structured data may become effectively invisible to machine systems. This transition may render legacy data portals, designed for manual browsing and human interpretation, obsolete, as machines require high-quality metadata, structured formats, and stable, authenticated access channels to remain functional.

AI systems are also becoming intermediaries between people and public information. Citizens increasingly ask AI platforms questions about public services, rights, or government procedures instead of consulting official sources directly. When these systems draw from unofficial, outdated, or incomplete information, they can produce inaccurate answers that public institutions have limited ability to see, correct, or contextualize. This creates new challenges around data quality, misinformation, unverifiable outputs, and the ability of public institutions to maintain authoritative information channels in a machine-mediated environment.

\subsection{Signal 3: Inference Is Reshaping the Foundations of Data Governance}
Several examples discussed during the studios pointed to another major challenge: valuable or sensitive knowledge can now be generated without direct access to the underlying data. Data governance has traditionally focused on collection, ownership, publication, and access---who holds a dataset, who can use it, and under what conditions. AI-enabled inference complicates that model because sensitive insights can increasingly be produced without any formal transfer or release of the underlying data.

One example involved a navigation application in Taiwan that displayed accurate red-light countdowns by aggregating vehicle stop patterns and GPS traces, even though the official traffic-light data was stored on protected government networks and had never been released publicly. The company inferred the rhythms of the city's infrastructure without ever obtaining access to the underlying dataset. The example illustrates how platforms can derive knowledge about public infrastructure, private behavior, or social systems from secondary traces, creating governance concerns even when no official data sharing has occurred.

As a result, governance questions are expanding beyond ownership and control of datasets themselves. Personal, anonymized, behavioral, environmental, and community-level data may generate sensitive downstream inferences when combined at scale, even if individual datasets appear harmless in isolation. The central question is increasingly shifting from who controls a dataset to who is affected by inferred knowledge, who can audit or challenge it, and what recourse exists when it produces real-world consequences.

Ownership still matters for identifying who holds an asset and for managing institutional risk, procurement, and accountability. Yet AI-driven inference requires additional governance tools focused on downstream impacts, provenance, contestability, reciprocity, and equitable value-sharing. As AI systems become more capable, data governance will increasingly involve managing the rights to sense, infer, and act on knowledge, reflecting an understanding of data as relational, generative, and produced through interactions rather than simply as a collection of static assets.

\subsection{Signal 4: Data Infrastructure Is Becoming Harder to Sustain}
The discussions also highlighted growing concerns about the future of data infrastructure, which is becoming a major constraint on whether information can remain accessible, usable, and trustworthy over time. AI-driven scraping, bot traffic, larger and more complex datasets, aging systems, and rising maintenance demands are placing increasing pressure on institutions responsible for maintaining public data infrastructure. Repositories and portals originally designed for human browsing are increasingly strained by machine-scale demand, while automated traffic can rapidly consume additional hosting, security, and bandwidth capacity without producing equivalent public value.

Maintaining data access is increasingly a resilience challenge. Long-term data availability can no longer be assumed in a landscape marked by cyberattacks, democratic backsliding, funding cuts, market consolidation, and threats to physical infrastructure such as undersea cables. In many contexts, basic reliability remains a foundational barrier, as power-grid instability, weak cybersecurity capacity, and under-resourced public systems limit the ability to maintain accessible and trustworthy data ecosystems.

The economics of openness are also becoming harder to sustain. Organizations face growing costs not only to publish new data, but to maintain, secure, document, authenticate, and serve existing datasets over time. This creates a growing ``who pays?'' dilemma. As democracy funders pull back and security, hosting, and quality-control costs rise, the long-term sustainability of open-data infrastructure is becoming increasingly uncertain.

Resilience creates additional trade-offs. Public-interest data may require backup systems, distributed preservation models, and redundancy strategies to guard against cyberattacks, political disruption, or institutional collapse. Yet large-scale duplication, cloud storage, and AI infrastructure carry significant financial and environmental costs, especially as data centers and AI workloads consume increasing energy and computing capacity. This creates pressure for more frugal infrastructure models and more selective preservation strategies. Infrastructure choices are also becoming more political: cloud dependence, jurisdictional exposure, and geopolitical rivalry are increasingly shaping decisions about where data is stored, who can access it, and under what conditions, prompting some institutions to reconsider access to sensitive infrastructure data where openness may create security or adversarial risks.

\subsection{Signal 5: Data Governance Is Fragmenting Across Institutions and Jurisdictions}
A recurring concern from both studios was the growing fragmentation of the data governance landscape. Within countries, different levels of government and agencies often pursue separate approaches to data access, privacy, AI, cybersecurity, digital infrastructure, and public-service delivery despite relying on the same underlying data ecosystems. Internationally, national regulations, regional frameworks, UN processes, development initiatives, and private-sector standards are evolving simultaneously, often with different assumptions, timelines, institutions, and definitions of risk. The result is a governance environment that is increasingly difficult for governments, businesses, and civil-society organizations to navigate.

Fragmentation is also visible across sectors. Health data, geospatial data, agricultural data, infrastructure data, financial data, and language data often require distinct institutional arrangements, governance models, and risk-management approaches. Local variation can be valuable when it reflects different cultural norms, development priorities, institutional capacities, and political realities. The challenge is to build enough interoperability, coordination, and institutional trust for diverse governance systems to function together without producing paralysis or conflict.

At the same time, new forms of governance infrastructure are emerging to bridge these divides. Digital economy agreements and digital trade agreements are increasingly operationalizing data governance across borders by creating mechanisms for interoperability, data exchange, and operational trust. These efforts reflect a shift toward building technical and institutional bridges that allow countries to maintain distinct legal frameworks while participating in shared digital economies.

Yet fragmentation remains a major implementation challenge, particularly for institutions with limited technical, legal, financial, or regulatory capacity. Participants also highlighted a growing gap between policy design and operational reality. Data governance is often driven by legal and compliance expertise without sufficient integration of technical, operational, and sociotechnical perspectives, producing frameworks that can be difficult to implement, audit, or translate into practice. Bridging this gap will require stronger stewardship capacity, multidisciplinary teams, technical translators, and institutional functions capable of connecting governance principles to the realities of how data systems are built, maintained, and used.

\subsection{Signal 6: Sovereignty and Security Are Driving a Turn Toward Strategic Control}
Participants spent considerable time discussing how geopolitical competition, security concerns, and dependence on foreign digital infrastructure are reshaping data governance. Assumptions that cross-border data flows are inherently beneficial are giving way to concerns about adversarial access, cloud dependence, infrastructure exposure, and control over strategically valuable data resources. Underlying many of these discussions was a concern about the weaponization of digital dependencies. Reliance on foreign cloud infrastructure, platforms, foundation models, semiconductors, software ecosystems, and connectivity networks is increasingly shaping how governments assess data policy, sovereignty, and risk. These dependencies can become sources of leverage during geopolitical conflict or economic rivalry, creating incentives for governments to pursue greater autonomy throughout the digital stack.

Governments and institutions are increasingly viewing unrestricted data flows through a national-security lens, particularly when high-quality public datasets---such as those tied to national resilience or critical systems including geospatial, infrastructure, health, and legal records---may reveal sensitive information. Participants noted that, in many contexts, traditional open-data policies remain formally in place even as the political will to support them has already evaporated in favor of restricting access and protecting digital borders.

Formulations such as ``open but local'' and ``interoperable but separate'' capture this emerging direction. Controlled-access systems, localization requirements, sovereign-cloud strategies, and sovereign-AI initiatives are gaining traction as governments seek to reduce external dependence, limit adversarial access, and retain greater control over strategically valuable data assets. These approaches point toward a future in which access remains possible within trusted environments, jurisdictions, or data spaces, while global openness becomes more limited.

Sovereignty is therefore becoming a dominant frame, but its meaning and feasibility remain contested. In some contexts, sovereignty reflects a legitimate effort to reduce dependency and strengthen local control over critical data systems. In others, it can become a reactive response that prioritizes restriction before addressing deeper questions of governance, capability, incentives, and public value. Participants repeatedly highlighted the tension between sovereignty as a political objective and sovereignty as an operational capability. Many governments continue to depend heavily on foreign cloud providers, AI systems, platform infrastructure, and technical services even while pursuing greater autonomy, creating difficult trade-offs between openness, interoperability, development priorities, economic dependence, and national control.

In the Global South, these concerns are often linked to fears of digital colonialism. Countries and communities worry that local data, language resources, community knowledge, and public-sector information may be extracted to build AI services that are later sold back to them. Data localization and strategic control are increasingly being pursued as responses, but participants emphasized that meaningful autonomy requires more than restricting access: it also depends on technical capacity, sustainable infrastructure, enforceable governance, and local value creation. In many regions, sovereignty is further complicated by the fact that large volumes of personal data have already leaked, circulated, or been collected through everyday digital interactions, making control harder to reclaim.

The discussions suggested that data sharing is increasingly contingent on control. Governments, communities, workers, and data holders are more willing to share when they can define conditions, monitor use, and rely on credible safeguards against misuse. Visibility, recourse, and benefit-sharing are becoming important foundations for trust in a machine-mediated data ecosystem.

\subsection{Signal 7: Data-Sharing Models Need Stronger Incentives and Benefit-Sharing Mechanisms}
Participants repeatedly pointed to a gap between the promise and reality of data sharing. Despite significant public investment, cross-sectoral data sharing has often produced fewer durable business models, public-value use cases, or institutional arrangements than expected. Governments, international organizations, and researchers have promoted data collaboratives, data spaces, marketplaces, trusts, exchanges, and interoperability frameworks, yet many continue to face persistent barriers, including weak incentives, unclear value propositions, legal uncertainty, technical complexity, high transaction costs, and uneven trust. Organizations that control valuable data may see little reason to share when risks are immediate and benefits are uncertain or captured elsewhere.

A related challenge is the absence of widely accepted frameworks for valuing data as an economic and strategic asset. There is little agreement on how to measure, compensate, govern, or distribute the value generated from non-personal, public-interest, and community data. As a result, institutions often struggle to distinguish responsible reuse from extraction, while communities and data holders may see little return from the value their data helps create. Uncertainty around reciprocity, benefit-sharing, and public value can make voluntary sharing less attractive and contributes to growing pressure to revisit existing data-sharing arrangements.

Participants argued that stronger enabling conditions are needed if data sharing is to scale. New institutional roles---including data stewards, technical mediators, intrasystemic mediators, and multidisciplinary trust brokers---can help negotiate access conditions, clarify accountability, connect data supply to public demand, and identify systemic risks across social, economic, and technical dimensions \citep{verhulst2020stewards,verhulst2019leveraging}. They can also help design mechanisms that make participation worthwhile and legitimate, including reciprocity requirements, community benefit agreements, shared-value models, local data trusts, licensing conditions, and public-return obligations. Yet they will face many of the same challenges that limited earlier sharing models unless they are supported by clear mandates, sustainable resources, enforceable governance, and credible mechanisms for distributing value and managing risk.

\section{Implications: Toward Anticipatory Data Governance}
\label{sec:implications}
Across the two studios, participants described a data-policy landscape undergoing profound change. Generative AI, machine-scale data reuse, geopolitical competition, infrastructure pressures, and growing concerns about extraction are challenging many of the assumptions that shaped open-data policy over the past two decades. Questions of access, trust, stewardship, sovereignty, infrastructure, and public value are becoming increasingly intertwined.

The seven signals suggest that the future of data governance will require new forms of stewardship, stronger interoperability mechanisms, more sustainable infrastructure, clearer benefit-sharing arrangements, and governance models capable of operating in complex and machine-mediated data ecosystems. While openness remains a foundational principle, the conditions under which data is shared, reused, and governed are being redefined.

The discussions also highlighted the importance of anticipatory governance. Forecasting specific technological breakthroughs may be difficult, but many of the pressures reshaping data ecosystems are already visible. Governments, institutions, and communities can prepare by strengthening their ability to identify emerging shifts, understand their implications, and adapt before risks, dependencies, or missed opportunities become embedded \citep{fuerth2009,maffei2020}. This requires systematic signal monitoring, multidisciplinary expertise, stronger feedback loops with affected communities, and governance processes that can evolve alongside technologies, incentives, and risks \citep{marcucci2025}.

\subsection{Reinforcing feedback loops}
Several of the signals reinforce one another, and treating them in isolation understates the dynamics at work. Machine-scale reuse increases demand for public-interest data, which can heighten concerns about extraction and push institutions toward more controlled access. Greater control can support trust and enable sharing, but it can also deepen fragmentation and increase the need for interoperability mechanisms. Infrastructure strain can make openness more expensive, which may narrow access unless sustainable funding and stewardship models are developed. Sovereignty concerns can prompt action that protects against dependency and adversarial use, but they can also accelerate localization, reduce cross-border collaboration, and make benefit-sharing harder to operationalize. These feedback loops indicate that interventions in one area may amplify risks or opportunities elsewhere in the data ecosystem, and that governance responses should be evaluated for their system-level effects rather than in isolation.

\subsection{Cross-cutting distinctions}
Several themes cut across the signals. Participants repeatedly emphasized the need to distinguish beneficial reuse from extraction, interoperability from dependency, sovereignty from closure, and openness from exposure. They also highlighted the growing importance of trust infrastructure---including provenance, authenticity, auditability, verification, recourse, and benefit-sharing mechanisms---that can support participation in increasingly machine-mediated environments.

\section{Conclusion}
\label{sec:conclusion}
Ultimately, the discussions suggest that data governance is a key foundation for AI governance, digital public infrastructure, democratic resilience, economic strategy, and public trust. As data ecosystems become more machine-mediated, interconnected, and contested, practitioners and policymakers will need to move beyond the challenge of making data available to also address how its use can generate public value while remaining trustworthy, sustainable, and aligned with societal goals. The opportunity ahead is to redesign the conditions under which data can continue to generate public value in a rapidly changing technological and geopolitical landscape. The seven signals presented here are offered not as predictions but as an evidence-informed framework for the anticipatory reasoning that this task will require.

\appendix
\section{Studio Participants}
\label{app:participants}
\begin{itemize}[leftmargin=1.4em,itemsep=0.15em]
\item \textbf{Audrey Tang} --- Cyber Ambassador and 1st Digital Minister, Government of Taiwan
\item \textbf{David Dab} --- Director of Digital Policy and Trust, Microsoft 365
\item \textbf{Edward Ghafari} --- CEO, iCES Corporation
\item \textbf{Gabriella Razzano} --- Executive Director, OpenUp
\item \textbf{Haishan Fu} --- Chief Statistician of the World Bank Group and Director of the Development Data Group, Development Economics (DEC)
\item \textbf{Hilde Hardeman} --- Director-General, Publications Office of the European Union
\item \textbf{Jean Noe Landry} --- Director, P\^ole sur les donn\'ees climatiques et d'adaptation, IRIU
\item \textbf{John Wilbanks} --- Independent Researcher
\item \textbf{Jowil Mejia Plecerda} --- Senior Officer, E-Commerce and Digital Trade, ASEAN Secretariat
\item \textbf{Leonida Mutuku} --- AI Research Director, Local Development Research Institute
\item \textbf{M\'elanie Dulong de Rosnay} --- Research Professor, National Centre for Scientific Research (CNRS)
\item \textbf{Michael Sch\"onstein} --- Head of General Digital Policy, German Federal Chancellery
\item \textbf{Monica Granados} --- Director of Open Science, Creative Commons
\item \textbf{Natalia Carfi} --- Executive Director, Open Data Charter
\item \textbf{Ot\'avio Moreira de Castro Neves} --- Head of Open Data Policy, Office of the Comptroller General (CGU), Government of Brazil
\item \textbf{Peter Rabley} --- CEO, Open Geospatial Consortium
\item \textbf{Steve MacFeely} --- Chief Statistician and Director of Statistics, OECD
\item \textbf{Tunde Fafunwa} --- Managing Partner, Kitskoo
\item \textbf{Yacine Jernite} --- Head of Machine Learning and Society, Hugging Face
\end{itemize}

\bibliographystyle{plainnat}
\bibliography{refs}

\end{document}